\documentstyle[twocolumn,aps,epsf]{revtex}

\begin{document}
\draft
\preprint{HEP/123-qed}
\title{Floppy Membranes}
\author{M. C. Diamantini\cite{bylinea}, H. Kleinert and C. A. Trugenberger}
\address{Institut f\"ur Theoretische Physik, Freie Universit\"at Berlin,\\
Arnimallee 14, D-1000 Berlin 33, Germany}
\date{\today}
\maketitle
\begin{abstract}
Floppy membranes are tensionless surfaces
without extrinsic stiffness whose
fluctuations are governed by
fourth-order bending elasticity. This
suppresses spiky superstructures and ensures that
floppy membranes
remain smooth
over any distance,
with
Hausdorff
dimension $D_H=2$,
in contrast to
surfaces with stiffness which are rough on the scale of
some finite persistence length.
\end{abstract}

\narrowtext
{\bf 1.} Under deformations, fluid membranes
behave approximately
like ideal tensionless surfaces with
curvarture stiffness \cite{helfrich,reviewd},
and have a model energy
\begin{equation}
H =\kappa \int d^2\xi \ \sqrt{g}
\ {\rm Tr} \ C^2       ,
\label{@hham}\end{equation}
where $C_{ab}$ is the second fundamental form of the surface
described by an embedding function ${\bf x}={\bf x} \left( \xi^1, \xi^a2 \right)$.
The symbol $g$ denotes the determinant of the induced
metric
 $g_{ab}\equiv \partial_a {\bf x} \partial _b{\bf x}$.

Thermal fluctuations are known to soften
the
curvature stiffness with increasing membrane size,
such that there exists
a finite persistence length
$\zeta = \zeta_0 \ {\rm exp}(4\pi \kappa/3T)$, 
where $\zeta_0$ is the molecular size,
beyond which the membrane looses its
stiffness completely
and begins exhibiting
a surface tension proportional to $\zeta^{-2}$, which was initially absent.
Thus the tangential correlation functions have only
the range $\zeta$,
and  surfaces much larger than $ \zeta $ appear rough.
The typical behavior of bilipid vesicles
can therefore be observed in the laboratory
only for sizes smaller than $\zeta$.
Surfaces much larger than $ \zeta $ will crumple and fill the
embedding space with spiky structures \cite{david,spiky}.

Apart from these common membranes, nature may
provide us also with
another type, not described by
the Hamiltonian (\ref{@hham}).
If the molecules in a bilayer are strongly conical,
the bending stiffness can be zero or negative  \cite{petrov}.
Such a situation can also arise
for
charged or dipolar molecules \cite{dipoles}.
In this case we shall speak of
{\em floppy membranes\/}.
The purpose of this note
is to study the statistical properties of such floppy membranes,
which will turn out to be quite different from those
of ordinary membranes.
In particular we shall find that, 
in contrast to ordinary membranes, floppy membranes without tension and stiffness
are
smooth
over long distances.
If the stiffness is negative,
floppy membranes are able to form
{\it disordered superstructures} similar to those
recently
observed in the laboratory \cite{reviewh}.
These are thought to be molten versions of
the egg carton-like
crystalline arrangement of local maxima and minima on the surface.
This has not yet been confirmed experimentally, but was suggested
by recent
numerical simulations \cite{egg}.

{\bf 2.} In order to describe floppy membranes,
we
must stabilize their fluctuations
 by adding to the
energy (\ref{@hham}) a higher-gradient
term. Here we shall
consider only one of 
several possibilities, focusing our attention upon
 the following Hamiltonian:
\begin{eqnarray}
H &&= r \int d^2\xi \sqrt{g} +{\kappa \over 2} \int d^2\xi \ \sqrt{g}
\ {\rm Tr} \ C^2
\nonumber \\
&&+{1\over 2m} \int d^2\xi \sqrt{g} \ \left[ g^{cd}{\cal D}_a C_{ac}
{\cal D}_b C_{bd} + {\rm Tr} \ C^4\right] \ .
\label{none}
\end{eqnarray}
The properties of a surface with this Hamiltonian will
be studied {\em non-perturbatively\/} in the limit 
of a large number $D$ of embedding dimensions.

The first term in (\ref{none}) parametrizes the surface tension; 
the third term
provides the surface with the stabilizing higher-order
bending stiffness, whose parameter
$m$
has the dimension $({\rm energy} \times  {\rm surface})^{-1}$.
Although this term
is irrelevant in a perturbative
renormalization group ana\-lysis,
it becomes relevant non-perturbatively in the limit
$D\rightarrow \infty$
 by a mechanism
familiar from
the three-dimensional Gross-Neveu model \cite{gross}.
The new term stabilizes the surface
against growing spikes and makes it smooth over long distances.

The Hamiltonian (\ref{none}) can be reformulated alternatively
in terms of the
tangent vectors
$\partial _a {\bf x}$ of the surface,
or in terms of
the normal
vectors ${\bf n}$:
\begin{eqnarray}
H &&= r \int d^2\xi \sqrt{g} -{1\over m }\int d^2\xi \ \sqrt{g} \ K^2
\nonumber \\
&&+{1\over 2}
\int d^2\xi \sqrt{g} \ {\bf n} \left[ -\kappa {\cal D}^2+{1\over m }
{\cal D}^4 \right] {\bf n}\ ,
\label{ntwo}
\end{eqnarray}
where  $K$ is the Gaussian curvature.
This form
exposes an important physical aspect of the model. As pointed
out in \cite{polyakov}, the
first term in the second line
is analogous to the
the
continuous version of the
Heisenberg model of ferromagnets \cite{zinn},
albeit with an additional integrability condition for the ${\bf
n}(\xi)$-field.
It tries to make an ordinary membrane with a positive stiffness
smooth
corresponding to a
ferromagnetic alignement of the normal vectors.
The fact that ordinary membranes cannot be smooth over
long distances
has its parallel in the absence of an ordered phase
in the
two-dimensional Heisenberg ferromagnet.
By the same analogy,
we see that
our new term introduces (apart from an intrinsic term $K^2$)
an antiferromagnetic next-to-nearest neighbours interaction between normal
vectors.
This generates frustration, and it is due to this
non-local
interaction that the surface can have an ordered phase after all, although
with an antiferromagnetic type of order.

To exhibit these features analytically,
a formulation of the Hamiltonian
(\ref{none}) in terms of the
tangent vectors will be most convenient:
\begin{equation}
H = {1\over 2} \int d^2{\xi } \sqrt{g} \ g^{ab}
{\cal D}_a x_{\mu }\! \left( r - \kappa {\cal D}^2 +
{1\over m} {\cal D}^4 \right)
\ {\cal D}_b x_{\mu } \ .
\label{nthree}
\end{equation}

{\bf 3.} We analyze the model (\ref{nthree}) in the
large-$D$ approximation
along the lines of Refs. \cite{david,klei}.
To this end we introduce a
Lagrange multiplier matrix $\lambda ^{ab}$
to enforce the constraint $g_{ab}=\partial _a{\bf x}\partial_b{\bf x}$,
extending the Hamiltonian (\ref{nthree}) to
\begin{equation}
H_{\rm ext}= H+ {1\over 2} \int d^2\xi \sqrt{g} \ \ \lambda
^{ab} \left( \partial _a {\bf x} \partial _b{\bf x} - g_{ab} \right) \ .
\label{nfour}
\end{equation}
Then we parametrize the surface in a Gauss map by
${\bf x} (\xi ) = \left( \xi _1, \xi _2, \phi ^i (\xi )
\right) $, $(i=3, \dots , D)$,
where $-R_1 /2\le \xi_1 \le R_1/2$,
$-R_2/2 \le \xi _2 \le R_2/2$ and
$\phi ^i(\xi )$ describe the ($D$-2)
transverse fluctuations.
In the limit of large $D$, to be studied here,
the large number of components suppresses the
fluctuations of $\lambda ^{ab}$ and $g_{ab}$.
These fields take extremal values which, for large
surface areas, are homogeneous and istropic:
$g_{ab}=\rho \ \delta_{ab} $,  $\lambda ^{ab} =
\lambda \ g^{ab}$.
Thus we may replace (\ref{nfour}) for large $D$ by
\begin{eqnarray}
H_{\rm ext}  &&= \int d^2\xi \ \left[ r +\lambda
(1-\rho ) \right]
\nonumber \\
&&+ {1\over 2} \int d^2\xi  \ \partial_a\phi ^i
\ K\left( -{\cal D}^2
\right) \ \partial_a \phi ^i\ ,
\label{nfive}
\end{eqnarray}
where $K$ represents the differential operator
\begin{eqnarray}
K\left(- {\cal D}^2 \right)
&&= r+\lambda  - \kappa {\cal D}^2 +
{1\over m} {\cal D}^4 \ .
\label{nsix}
\end{eqnarray}
Integrating out the transverse fluctuations, always for
large areas, we get the free energy
\begin{eqnarray}
F &&= A_{\rm ext} \ \left[ r+\lambda
(1-\rho ) \right]
\nonumber \\
&&+ A_{\rm ext} {{D-2}\over 8\pi^2 }\rho
\int d^2p\ {\rm ln}
\left[ p^2 K\left( p^2 \right)
\right] \ ,
\label{nseven}
\end{eqnarray}
where, for simplicity, we have chosen natural units
by setting $\beta =1/k_{\rm B}T=1$ and
$A_{\rm ext}=R_1 R_2$ is the
extrinsic, projected area in the coordinate plane.
The factor $(D-2)$ in the second term ensures that,
for large $D$, the fields $\lambda$ and $\rho$ are extremal
and satisfy thus the saddle-point (``gap") equations
\begin{equation}
0 = f\left( r, \kappa, m, \lambda \right) \ ,\quad
\rho = {1\over f'\left( r, \kappa, m , \lambda \right) }\ .
\label{neight}
\end{equation}
The prime denotes derivatives with respect to $\lambda$ and
the saddle-point function $f$ is defined by
\begin{equation}
f\left( r, \kappa, m, \lambda \right) \equiv \lambda
- {{D-2}\over {8\pi }}
\int dp \ p\  {\rm ln}
\left[ p^2 K
\left( p^2 \right) \right] .
\label{nove}
\end{equation}
Inserting (\ref{neight}) in (\ref{nseven}) we get
$F= \left( r+\lambda \right)
\ A_{\rm ext}$
showing that $r_{\rm ph} \equiv \left( r+\lambda \right)$
is the physical surface tension.

We now introduce the following combinations of the parameters
of the model:
\begin{equation}
R^2 \equiv {1\over 2} \sqrt{m (r+\lambda)}
+ {\kappa m\over 4}\ ,\quad
I^2 \equiv {1\over 2} \sqrt{m (r+\Lambda )}
- {\kappa m\over 4}\ .
\label{nten}
\end{equation}
In terms of these, the kernel $V$ can be written as
\begin{equation}
m K\left( p^2 \right) =
\left( R^2+I^2 \right) ^2 + 2 \left( R^2-I^2 \right) p^2
+ p^4 \ ,
\label{neleven}
\end{equation}
from where we deduce the stability condition for the homogeneous
saddle-point as being $R^2>0$, insuring that the spectrum of transverse
fluctuations is positive for all $p>0$.
Since we shall mostly consider negative
or vanishing stiffnesses $\kappa $
we shall also assume that $I^2>0$, so that both
$R$ and $I$ are real.
Note that the spectrum of transverse
fluctuations depends drastically on the sign of the stiffness: for negative
$\kappa $ we have $I>R$ and $K$ develops a minimum at
$p=\sqrt{I^2-R^2}$. Correspondingly (but for a slightly higher value
of $I/R$), the spectrum $E\left( p^2 \right) =
p^2 K\left( p^2 \right)$
develops a roton-like minimum, which, for $R/I \ll 1$,
lies at $p\simeq I\left( 1-8R^2/3I^2 \right)$,
as shown schematically in Fig. 1.

\begin{figure}
\centerline{\epsfysize=5.5cm\epsfbox{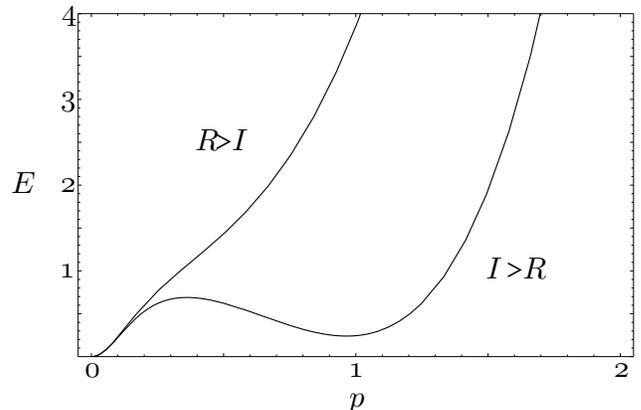}}
\caption{The roton-like minimum in the spectrum of transverse fluctuations
for $I\gg R$.}
\end{figure}

Having established the stability conditions we proceed to the evaluation of
the saddle-function. This contains
an ultraviolet divergent integral which must be
regularized. To this end we use the standard
dimensional regularization, computing
the integral in $(2-\epsilon)$ dimensions.
For small $\epsilon$, this leads to
\begin{eqnarray}
f\left( r, \kappa, m, \lambda \right) &&= \lambda
+ {1\over 4 \pi } \left( R^2-I^2
\right) {\rm ln} {R^2+I^2\over \Lambda ^2} \\
\nonumber
&& - {1\over 2 \pi } RI \left( {\pi \over 2} + {\rm arctan}
{I^2-R^2 \over 2RI} \right) \ ,
\label{insone}
\end{eqnarray}
where $\Lambda \equiv \mu \ {\rm exp} (2/\epsilon)$ and $\mu $ is a
reference scale which must be introduced for dimensional reasons.
The scale $\Lambda $ plays the role of an
ultraviolet cutoff,
diverging for $\epsilon \to 0$.

In order to distinguish the various phases of our model we
compute two correlation functions. First, we
consider the orientational correlation function
$g_{ab}(\xi -\xi ') \equiv \langle \partial_a \phi ^i (\xi )
\ \partial_b \phi ^i (\xi ') \rangle $ for the normal components of
tangent vectors to the surface. From (\ref{nfive}) this is given by
\begin{equation}
g_{ab}(\xi -\xi ') = {\delta _{ab} \over 4\pi ^2}
\ \int d^2p \ {1\over
K\left( p^2 \right) }
\ \ {\rm e}^{i \sqrt{\rho } p (\xi -\xi ')} \ .
\label{ntwelve}
\end{equation}
In terms of $R$ and $I$, the Fourier
components can be writtens as
\begin{equation}
{1\over K\left( p^2 \right) } =
{m\over 2RI} \ {\rm Im} \ {1\over {p^2 + (R-iI)^2}} \ ,
\label{thirteen}
\end{equation}
from where we obtain the following exact result for the diagonal
elements $g  \equiv g_{aa}$ of (\ref{ntwelve}):
\begin{equation}
2 \pi  g (d) =
{m\over 2RI}
\ {\rm Im}\ K_0\left( \left( R-iI \right)
\sqrt{\rho} d \right) \ ,
\label{nfourteen}
\end{equation}
where $d\equiv |\xi -\xi'|$ and $K_0$ is a
Bessel function of imaginary argument
\cite{gr}.

Secondly, we compute the scaling law of
the distance $d_E$ in embedding space
between two points on the
surface  when changing its projection $d$
on the reference plane. The exact relation between
the two lengths is
\begin{equation}
d_E^2=d^2 + \sum_i \langle |\phi ^i (\xi) - \phi ^i (\xi ')
|^2\rangle \ .
\label{nfifteen}
\end{equation}
With a computation analogous to the one leading to (\ref{nfourteen})
we obtain the following behaviour:
\begin{eqnarray} \nonumber
d_E^2= \cases{\displaystyle{
\left( R^2+I^2 \right) \over 8 \pi r_{\rm ph} RI}
 {\rm arctan} (I/R)  \alpha d^2, &$\!\!\!d^2 \ll
{1\over \alpha}$ \cr\cr
\displaystyle
{1\over 2\pi r_{\rm ph}}  \left[ {\rm ln}\left( \alpha
d^2 /4 \right) + c(R, I) \right] ,
&$\!\!\!{1\over \alpha}\ll d^2
\ll {1\over 2\pi r_{\rm ph}}$ \cr \cr
\displaystyle
\,d^2, &$\!\!\!\!d^2\gg {1\over 2\pi r_{\rm ph}}$ 
}
\nonumber
\end{eqnarray}
where $\alpha \equiv \left( R^2+I^2 \right) \rho $ and $c(R, I)=
C+ \left[\left(I^2-R^2 \right) /RI \right] {\rm arctan} (I/R)$
with $C$ = Euler's constant.

These results show that the model has three possible phases.
The first is realized when there are no solutions to the saddle-point
equations in the allowed range of parameters.
For this choice of parameters, there exist no
homogeneous, isotropic surfaces. In this phase the surfaces will
form inhomogeneous structures.

If a solution to the saddle-point equations exists, two situations can
be realized. For large positive stiffness $\kappa $ we have
$R\gg I$, the asymptotic region is $d
\gg 1/R\sqrt{\rho}$ and $I$ can be neglected.
In this region we have
\begin{equation}
g(d) \propto {1\over \sqrt{R\sqrt{\rho}d}} \ {\rm e}^
{-R\sqrt{\rho}d} \ ,
\label{nsixteen}
\end{equation}
exhibiting short-range orientational order.
For short distances $d \ll 1/R\sqrt{\rho}$,
the surfaces behave as two-dimensional
objects.
If $\rho$ becomes large we have a region $1/R\sqrt{\rho} \ll
d\ll 1/\sqrt{2\pi r_{\rm ph}}$ in which $d_E$ scales logarithmically with
$d$ and distances along the surface become large. The transition to this
regime happens on the scale of the persistence length
$d_E^{\rm PL}=1/\sqrt{r_{\rm ph}}$.
Above this scale world-sheets are crumpled,
with no orientational correlations (if the tension is not large enough
to dominate over the entire surface,
causing $\rho \simeq 1$). This phase
corresponds to the familiar behaviour of
stiff membranes \cite{david}.

For large negative stiffness $\kappa $, in
contrast, we have
$I\gg R$, the asymptotic region is $d
\gg 1/I\sqrt{\rho}$, and $R$ can be neglected in $K_0$ for $d
\ll 1/R\sqrt{\rho} $. In this region we have
\begin{equation}
g(d)= {m \over 8 RI}
\ J_0 \left( I\sqrt{\rho}d \right) \ ,
\label{nseventeen}
\end{equation}
with $1/R\sqrt{\rho} $ playing the role of
an infrared cutoff for the oscillations
on the scale $1/I\sqrt{\rho}$ over which the Bessel function
$J_0$ varies. We have thus a new scale (in embedding space)
\begin{equation}
d_E^O= \sqrt{I\over {32 R r_{\rm ph}}}\ ,
\label{neighteen}
\end{equation}
on which the transverse fluctuations create oscillations characterized
by the ``antiferromagnetic" orientational
correlations (\ref{nseventeen}). Crumpling takes place only if
$1/R\sqrt{\rho} \ll 1/\sqrt{2\pi r_{\rm ph}}$ and the corresponding
persistence length is now
\begin{equation}
d_E^{\rm PL}={1\over \sqrt{4\pi r_{\rm ph}}} \ \sqrt{{\pi I \over 2R} + {\rm
ln}
{I^2\over 4R^2}}\ ,
\label{nnineteen}
\end{equation}
which is much larger than $1/\sqrt{r_{\rm ph}}$.
Otherwise, the oscillating superstructure
goes over directly into the tension dominated region.
In this case $d_E$
scales logarithmically with $d$ for $1/I\sqrt{\rho} \ll d
\ll 1/\sqrt{2\pi r_{\rm ph}}$, and $\rho $ is large not because
of crumpling but because of the oscillating superstructure.
Indeed there are strong orientational correlations in
this region. Note that the oscillations represent a disordered
superstructure caused by fluctuations on an otherwise homogeneous
and isotropic ground-state described by the solution of the saddle-point
equations.

{\bf 4.} One might imagine that our disordered superstructure undergoes a
transition
to a crystalline egg-carton-type
structure \cite{egg} when $R\to 0$, so that the spectrum
of transverse fluctuations develops an instability at a finite value $p=I$.
However this is not so, as can be seen from the
explicit expression for $\rho $ obtained
from (\ref{neight}),
\begin{equation}
\rho = \left[ 1-{m\over 16 \pi RI} \left({\pi\over 2} +
{\rm arctan} {I^2-R^2\over 2RI} \right) \right] ^{-1} \ .
\label{ntwenty}
\end{equation}
When lowering $R$ at fixed $I$ one hits a pole where $\rho$ diverges.
This means that the surface crumples
before one reaches the crystal instability.

For symmetry reasons, the transition
from the stiff to the disordered superstructure
phase occurs on the line $R=I$, where $\kappa =0$ and the
the kernel $K$ develops its minimum.

The fourth-order bending elasticity term dominates the
fluctuations of surfaces which have both vanishing bare tension
and stiffness. These can be studied further analytically since,
for $\kappa =0$ ($R=I$), the saddle-point equations become polynomial,
with solution
\begin{equation}
\lambda = {m \over 128}\ \left(
1+\sqrt{1+ 256 {r\over m }} \right) \ .
\label{newfive}
\end{equation}
This gives
\begin{equation}
r_{\rm ph} = {a^2 \over 64} \ m\ ,\qquad
\rho=\left(1 -{1\over 2a}\right) ^{-1}\ ,
\label{uno}
\end{equation}
where $a$ is the following function of the dimensionless
parameter $r/m$:
\begin{equation}
a^2 = {{1+128 \ r/m +\sqrt{1+ 256 \ r/m}}\over 2} \ .
\label{due}
\end{equation}
This is the previously announced result. The fourth-order
bending elasticity term, although irrelevant in
perturbation theory, becomes relevant non-perturbatively by
contributing a term $m/64$ to the physical surface tension.
This is the reason why,
contrary to stiff membranes \cite{david}, {\it floppy membranes do not
crumple}. The physical tension can be decreased arbirarily
with
$\rho$ remaining in the range $1\le \rho \le 2$. In other words one can
lower the two scales $r$ and $m$ so that the range of
orientational correlations $d=1/R\sqrt{\rho} = (4/a\sqrt{m}) \sqrt{a-1/2}$
is always of the same order or larger
than the inverse of the square root of the
physical surface tension $1/\sqrt{r_{\rm ph}} = 8/a\sqrt{m}$.

The corresponding $\gamma $-functions are easily obtained as
\begin{eqnarray}
\gamma (r) &&\equiv -\Lambda \ {d\over d\Lambda }
\ {\rm ln} \ \frac{r}{\Lambda ^2} = 2\ ,\\
\gamma (m) &&\equiv -\Lambda \ {d\over d\Lambda }
\ {\rm ln} \ \frac{m}{\Lambda ^2 }= 2\ ,
\label{tre}
\end{eqnarray}
showing the absence of anomalous dimensions for $\kappa =0$.
Correspondingly, we have
\begin{equation}
\beta \left( {r\over m} \right)  = -\Lambda \ {d \over d\Lambda }
\ {r\over m}  = {r\over m}\left[ \gamma (r)-\gamma(m) \right] =0 \ ,
\label{quattro}
\end{equation}
which means that $r/m$, and thus also $a$, are renormalization
group invariants.
 
{\bf 5.} In conclusion we see that the physics of floppy membranes is
governed by an infrared-stable
fixed-point characterized
by vanishing tension and a dimensionless renormalization group invariant
parameter $a^* \equiv \lim_{r\to 0, m\to 0} a(r, m)$. 
At this point, the surface
exhibits long-range order in which the diagonal elements of the
correlation functions $g_{ab}(\xi-\xi')$
do not depend on the distance, $g(d) = 4\pi ^2/a^*$, and
in which the length (\ref{nfifteen}) scales
with  the distance in coordinate space like
 $d_E^2 = \left( \pi ^2 \rho ^*/a^* \right) d^2$,
from which we deduce
a Hausdorff dimension $D_H=2$
for floppy membranes.

\end{document}